\documentstyle[aps,prl,twocolumn,floats,epsf]{revtex}

\begin{document}



\title{ First  Observation of the  Decay  {\boldmath $B \to J/\psi  \,
\phi \, K$}} 
\author{
A.~Anastassov,$^{1}$ J.~E.~Duboscq,$^{1}$ K.~K.~Gan,$^{1}$
C.~Gwon,$^{1}$ T.~Hart,$^{1}$ K.~Honscheid,$^{1}$
D.~Hufnagel,$^{1}$ H.~Kagan,$^{1}$ R.~Kass,$^{1}$
J.~Lorenc,$^{1}$
% ...
}

\author{(CLEO Collaboration)}

\address{   $^{1}${Ohio  State  University,   Columbus, Ohio  43210}\\
% ...
} \date{\today} \maketitle 

\begin{abstract} 
We present the first observation of the decay $B \to J/\psi \, \phi \,
K$.  Using $9.6\times10^6$ $B \overline B$  meson pairs collected with
the  CLEO detector,  we have observed  10  fully  reconstructed $B \to
J/\psi \,  \phi \, K$ candidates,  whereas the estimated background is
$0.5\pm0.2$ events.  We obtain a branching fraction of ${\cal B}(B \to
J/\psi  \,   \phi  \,   K)  =  (8.8^{+3.5}_{-3.0}[\rm  stat]\pm1.3[\rm
syst])\times  10^{-5}$.  This is the   first observed $B$ meson  decay
requiring the creation of an additional $s \overline s$ quark pair. 
\end{abstract}
\pacs{13.25.Hw} 

\begin{figure*}[htb]
\centering \epsfxsize=10cm \epsfysize=6cm \epsfbox{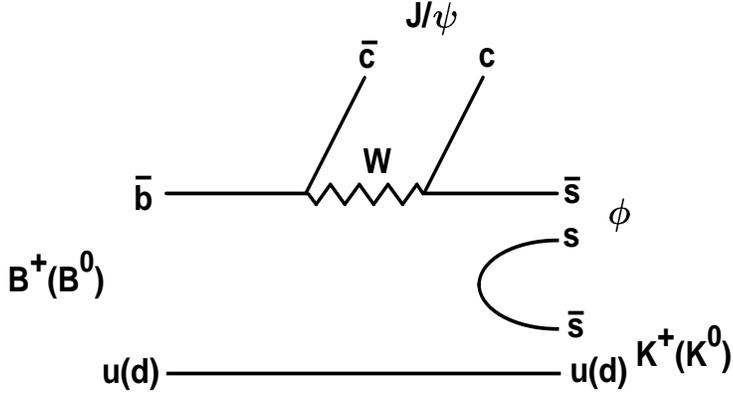} 
\caption{Most likely $B \to J/\psi \, \phi \, K$ decay mechanism.  } 
\label{fig:fd_b_psi_phi_k}
\end{figure*}

An observation of   a $B$ meson decay requiring   the creation of   an
additional $s \overline s$ quark pair in the final state would enhance
our  understanding of strong  interactions in the  final states of $B$
decays.  Previous studies of such  processes involved searches for the
``lower     vertex''     $\overline        B  \to         D^+_s     X$
transitions~\cite{lower_vertex}, however no  signal was observed.  The
decay $B  \to J/\psi \,  \phi  \, K$~\cite{charge-conjugate} can occur
only if an  additional $s \overline s$ quark   pair is created in  the
decay chain besides the quarks produced in the weak  $b\to c \bar c s$
transition. The $B \to J/\psi  \, \phi \,   K$ transition most  likely
proceeds         as  a                    three-body             decay
(Fig.~\ref{fig:fd_b_psi_phi_k}). Another  possibility  is  that the $B
\to J/\psi \, \phi \, K$  decay proceeds as  a quasi-two-body decay in
which the $J/\psi$  and  $\phi$  mesons  are  daughters of  a   hybrid
charmonium state~\cite{Close:1998wp}.

We searched for $B^+ \to J/\psi \, \phi \, K^+$ and $B^0 \to J/\psi \,
\phi \, K^0_S$ decays,    reconstructing $J/\psi \to  \ell^+  \ell^-$,
$\phi \to  K^+ K^-$, and $K^0_S  \to  \pi^+ \pi^-$. Both  $e^+e^-$ and
$\mu^+\mu^-$   modes were used  for the  $J/\psi$ reconstruction.  The
data were collected  at the Cornell  Electron Storage Ring (CESR) with
two    configurations      of   the      CLEO    detector,      called
CLEO~II~\cite{Kubota:1992ww} and  CLEO~II.V.    The components  of the
CLEO detector most relevant to this  analysis are the charged particle
tracking  system,  the    CsI    electromagnetic calorimeter,      the
time-of-flight system, and the muon chambers.  In CLEO~II, the momenta
of charged particles are measured in a tracking system consisting of a
6-layer  straw tube chamber,  10-layer   precision drift chamber,  and
51-layer main drift  chamber, all operating inside  a 1.5 T solenoidal
magnet.   The  main drift chamber  also  provides a measurement of the
specific ionization  loss, $dE/dx$,  used for particle identification.
For  CLEO~II.V, the    innermost  wire chamber  was   replaced  with a
three-layer    silicon    vertex  detector~\cite{Hill:1998ea}.    Muon
identification  system consists   of proportional counters  placed  at
various depths in the steel absorber. 

The results of this search are  based upon an integrated luminosity of
 9.1 $\rm fb^{-1}$ of $e^+e^-$ data taken at the $\Upsilon(4S)$ energy
 and  4.4 $\rm fb^{-1}$   recorded   60~MeV below  the  $\Upsilon(4S)$
 energy.  The  simulated event  samples   used in this   analysis were
 generated   using   GEANT-based~\cite{GEANT}  simulation  of the CLEO
 detector   response.  Simulated events   were processed  in a similar
 manner as the data. 

When making requirements on such kinematic variables as invariant mass
or    energy,   we  took advantage   of     well-understood  track and
photon-shower covariance matrices to calculate the expected resolution
for  each  combination.    Therefore  we extensively  used  normalized
variables, which allowed uniform  candidate  selection criteria to  be
used for the data  collected with the CLEO II  and CLEO II.V  detector
configurations. 
 
\begin{figure*}[htbp]
\centering \epsfxsize=16cm \epsfysize=7cm \epsfbox{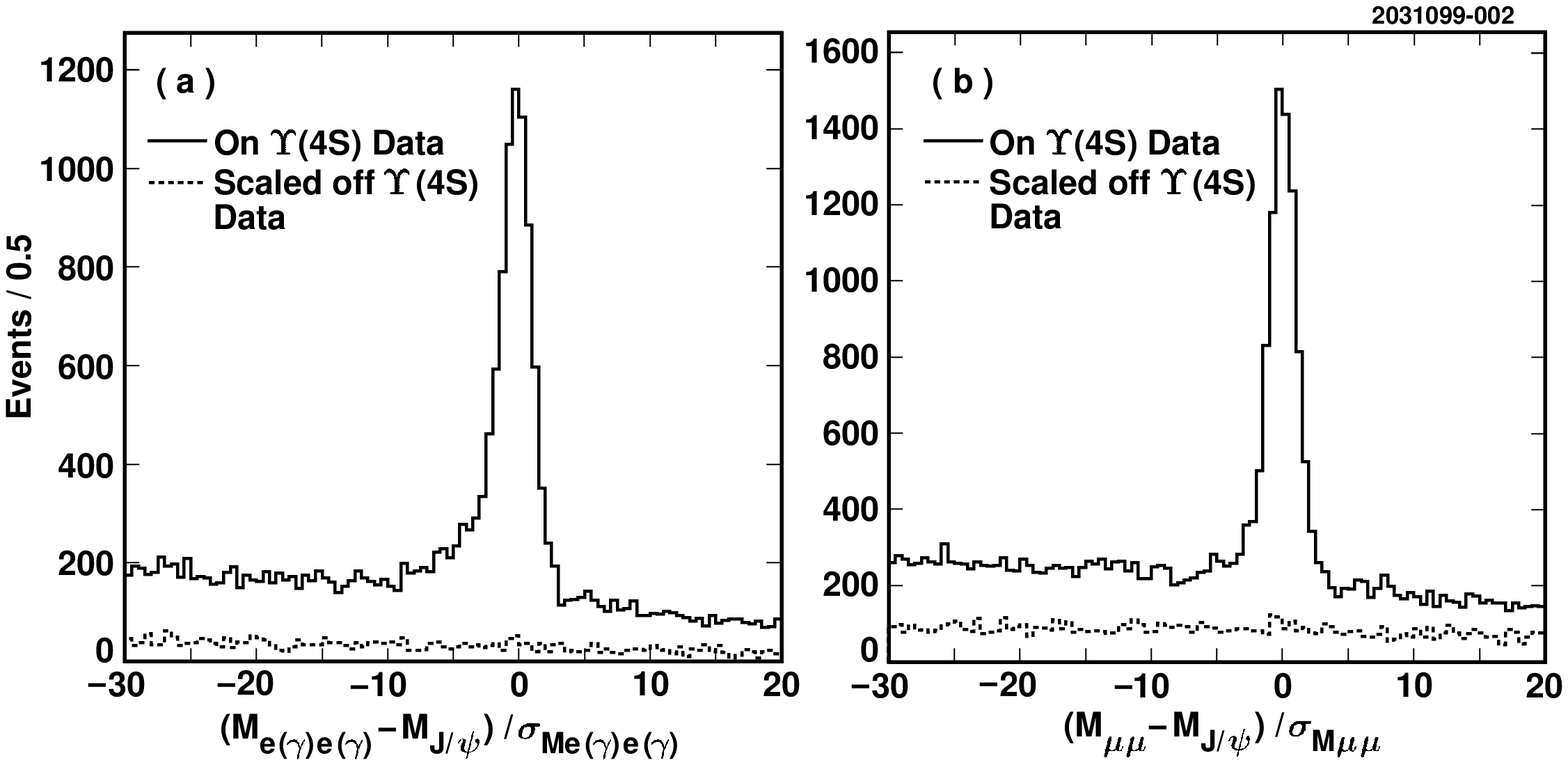} 
\caption{Normalized invariant mass of the (a) $J/\psi \to e^+ e^-$ and
(b) $J/\psi  \to \mu^+  \mu^-$ candidates  in  data.  The   solid line
represents the data taken  at  the $\Upsilon(4S)$ energy;  the  dashed
line represents the luminosity-scaled  off resonance data showing  the
level of background from non-$B \overline B$ events.  } 
\label{fig:data_ee_mumu_on_off}
\end{figure*}

The normalized invariant mass distributions for the $J/\psi \to \ell^+
\ell^-$     signal      in         data       are         shown     in
Fig.~\ref{fig:data_ee_mumu_on_off}.   We   required     the normalized
invariant mass  to be from $-10$  to $+3$ (from $-4$  to $+3$) for the
$J/\psi  \to  e^+  e^-$($J/\psi \to  \mu^+   \mu^-$) candidates.   The
resolution in the    $\ell^+  \ell^-$  invariant  mass is    about  10
MeV$/c^2$.   To  improve  the energy    and momentum resolution  of  a
$J/\psi$ candidate,  we performed a fit constraining  the mass of each
$J/\psi$ candidate to the world average value~\cite{PDG}.

 Electron  candidates were identified based on  the ratio of the track
momentum  to the associated shower energy   in the CsI calorimeter and
specific  ionization   loss   in the   drift  chamber.    The internal
bremsstrahlung  in the  $J/\psi  \to  e^+ e^-$ decay  as  well  as the
bremsstrahlung in the detector  material produce a long radiative tail
in the  $ e^+ e^-$   invariant mass distribution  and impede efficient
$J/\psi  \to   e^+  e^-$  detection.     We recovered  some    of  the
bremsstrahlung  photons by  selecting    the photon shower   with  the
smallest opening  angle with respect to  the direction of  the $e^\pm$
track evaluated at  the   interaction point, and  then  requiring this
opening angle  to  be  smaller  than $5^\circ$. The  addition  of  the
bremsstrahlung     photons  resulted  in    a    relative  increase of
approximately  25\%  in the   $J/\psi   \to e^+  e^-$   reconstruction
efficiency without adding more background. 

 For  the  $J/\psi \to \mu^+  \mu^-$  reconstruction, one  of the muon
candidates was  required to  penetrate the steel  absorber to  a depth
greater  than 3 nuclear  interaction lengths.  We relaxed the absorber
penetration  requirement for the  second muon candidate  if it was not
expected to reach a muon chamber either because its energy was too low
or because it pointed to  a region of the  detector not covered by the
muon  chambers.  For these  muon candidates we required the ionization
signature in the CsI calorimeter to be consistent with that of a muon.
Muons typically  leave a   narrow trail   of  ionization and   deposit
approximately 200~MeV of energy  in the crystal calorimeter.  Hadrons,
on the other hand,  quite often undergo a  nuclear interaction in  the
CsI crystals that  have  a  depth of   80\% of  a nuclear  interaction
length.  Compared to imposing  the absorber penetration requirement on
both muon  candidates, this procedure  increased the $J/\psi \to \mu^+
\mu^-$ reconstruction efficiency   by  20\%  with 80\% increase     of
background. 

We required  that  the charged kaon candidates  have  $dE/dx$ and,  if
available,  time-of-flight  measurements that   lie within  3 standard
deviations of the expected values. 

If for  the  $B \to J/\psi  \,  \phi\, K$ decays  we  assume a uniform
Dalitz   distribution and isotropic    decays  of $J/\psi$ and  $\phi$
mesons, then   the  expected efficiency of  the   combined $dE/dx$ and
time-of-flight selection  is  approximately 90\%  per  kaon candidate.
The $dE/dx$ measurements alone  provide the $K/\pi$ separation of more
than  4 standard deviations for 92\%  of the $\phi$ daughter kaons and
for 64\% of the ``bachelor'' kaons from $B$  decay.  We selected $\phi
\to K^+ K^-$ candidates  by requiring the  $K^+ K^-$ invariant mass to
be within  10~MeV/$c^2$ of the $\phi$  mass~\cite{PDG}. We did not use
the normalized $K^+ K^-$ invariant  mass  because the mass  resolution
(1.2~MeV/$c^2$) is smaller than the $\phi$ width (4.4~MeV)~\cite{PDG}.

The $K^0_S$  candidates were  selected from  pairs of  tracks  forming
well-measured displaced  vertices.  The resolution  in $\pi^+   \pi^-$
invariant mass is about 4~MeV$/c^2$. We required the absolute value of
the normalized $\pi^+ \pi^-$ invariant mass to be less than 4, then we
performed a fit constraining the mass of each $K^0_S$ candidate to the
world average value~\cite{PDG}. 

\begin{figure*}[htbp]
\centering \epsfxsize=16cm \epsfysize=7cm \epsfbox{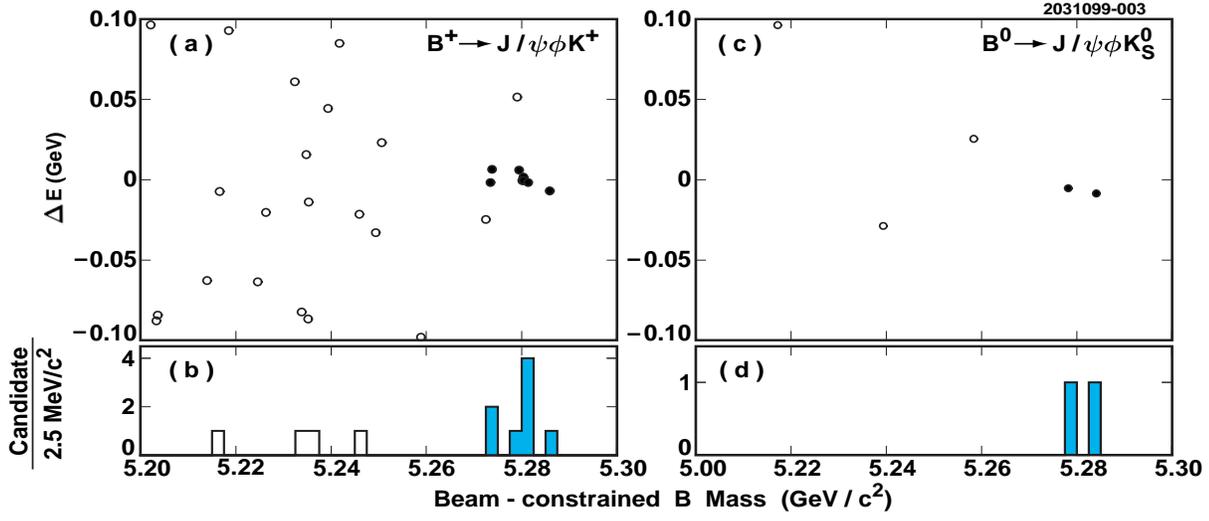} 
\caption{  The  $\Delta E$  vs  $M(B)$ distribution  for  (a) $B^+ \to
J/\psi   \, \phi \, K^+$  and  (c) $B^0  \to J/\psi \,  \phi \, K^0_S$
candidates in data.   The signal candidates, selected using normalized
$\Delta E$  and $M(B)$ variables,  are  shown by  filled circles.  The
$M(B)$ distribution  for (b) $B^+  \to J/\psi \, \phi  \, K^+$ and (d)
$B^0  \to  J/\psi \,  \phi \,   K^0_S$ candidates  satisfying $|\Delta
E/\sigma(\Delta E)|<3$; the shaded  parts of the  histograms represent
signal candidates. } 
\label{fig:nde_nmb_data}
 \end{figure*}

 The $B \to J/\psi \, \phi \, K$ candidates were  selected by means of
two  observables. The first  observable  is the difference between the
energy of  the $B$  candidate and the  beam  energy  $\Delta E  \equiv
E(J/\psi)+ E(\phi)+E(K) - E_{\rm beam}$.  The resolution in $\Delta E$
for the $B \to J/\psi \, \phi \, K$ candidates is approximately 6~MeV.
The   second    observable   is  the     beam-constrained  $B$    mass
$M(B)\equiv\sqrt{E^2_{\rm beam}-p^2(B)}$, where $p(B)$ is the absolute
value of the $B$ candidate momentum. The  resolution in $M(B)$ for the
$B \to J/\psi \, \phi \,  K$ candidates is  about 2.7~MeV/$c^2$; it is
dominated by the beam energy spread.  The distributions of the $\Delta
E$ vs $M(B)$ for  $B^+ \to J/\psi \, \phi  \, K^+$ and $B^0 \to J/\psi
\, \phi  \, K^0_S$ are  shown in Fig.~\ref{fig:nde_nmb_data}.  We used
the  normalized $\Delta E$ and $M(B)$  variables to  select the $B \to
J/\psi \,  \phi \,  K$ candidates and  defined   the signal region  as
$|\Delta E/\sigma(\Delta E)|<3$ and $|(M(B)-M_B)/\sigma(M(B))|<3$.  We
observed 8(2) events  in the signal region for  the $B^+ \to J/\psi \,
\phi  \, K^+$ ($B^0  \to J/\psi \, \phi \,  K^0_S$) mode.  Considering
that $K^0$  can decay as $K^0_S$  or as $K^0_L$,  and also taking into
account ${\cal  B}(K^0_S  \to  \pi^+  \pi^-)$ and  the   difference in
reconstruction efficiencies, we expect to observe  on average 4.3 $B^+
\to J/\psi \,  \phi \, K^+$ candidates   for every $B^0 \to  J/\psi \,
\phi \, K^0_S$ candidate. 

\begin{figure*}[htbp]
\centering \epsfxsize=7.0cm \epsfysize=7.0cm \epsfbox{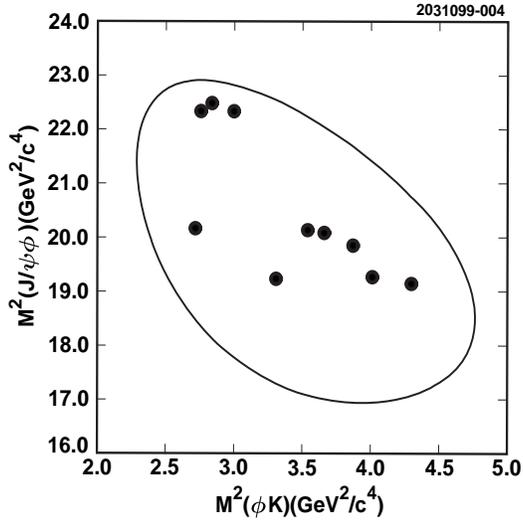} 
\caption{  Dalitz   plot for the  10  $B  \to J/\psi   \, \phi   \, K$
candidates  in data.  The kinematic boundary  is  represented by solid
line.} 
\label{fig:dalitz_data_contour} 
\end{figure*}

\begin{figure*}[htbp]
\centering \epsfxsize=7.0cm \epsfysize=7.0cm \epsfbox{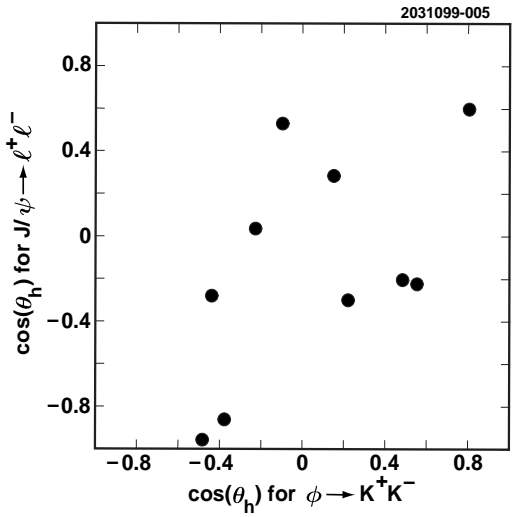} 
\caption{ The distribution of the cosine of helicity angle for $J/\psi
\to \ell^+ \ell^-$ vs  the cosine of helicity  angle for $\phi \to K^+
K^-$ for the 10 $B \to J/\psi \, \phi \, K$ candidates in data.} 
\label{fig:paper_polarization_data}
\end{figure*}

The Dalitz plot and the cosine of helicity angle distributions for the
10 $B  \to   J/\psi \, \phi \,   K$  signal candidates are    shown in
Figs.~\ref{fig:dalitz_data_contour}~and~\ref{fig:paper_polarization_data}.
The helicity angle for $J/\psi \to \ell^+ \ell^-$  decay is defined as
the angle between a lepton momentum in the $J/\psi$ rest frame and the
$J/\psi$ momentum in the $B$  rest frame. An analogous definition  was
used for the $\phi \to K^+ K^-$ decay.  No conclusion can be drawn yet
either about the  $J/\psi$ and the $\phi$  polarizations  or about the
resonant substructure of  the $B \to J/\psi \,  \phi \,  K$ decay.  If
the  $J/\psi$ and  $\phi$  mesons are  the   products  of  the  hybrid
charmonium $\psi_g$ decay, then the $J/\psi \, \phi$ invariant mass is
expected to  be below the  $DD^{**}$ threshold (4.3~GeV/$c^2$) because
$\psi_g    \to  DD^{**}$  decay is   likely    to  dominate above  the
threshold~\cite{Close:1998wp}.  The $J/\psi \, \phi$ invariant mass is
above 4.3~GeV/$c^2$ for all 10 $B \to J/\psi  \, \phi \, K$ candidates
thus difavoring the hybrid charmonium dominance scenario. 

The background can be divided into  two categories.  First category is
the combinatorial background from $\Upsilon(4S) \to B \overline B$ and
continuum    non-$B \overline  B$   events.    Second category is  the
background from non-resonant $B \to J/\psi \, K^+ \, K^- \, K$ decays. 

The  combinatorial  background from $\Upsilon(4S)   \to B \overline B$
events was estimated using a sample  of simulated events approximately
32 times the data sample; events containing a $B \to  J/\psi \, K^+ \,
K^-   \, K$ decay   were excluded.  We  estimated  the background from
$\Upsilon(4S)  \to B \overline  B$ decays to be $0.25^{+0.10}_{-0.08}$
events. In addition, we specifically considered $B \to J/\psi \, K^{*}
\, \pi^+$ with  $K^{*}\to K \pi^-$ and  $B \to J/\psi  \, \rho^0 \, K$
decays because the  beam-constrained  $B$ mass distribution  for these
modes  is  the same as for   the $B \to J/\psi \,   \phi \, K$ decays.
Using data and  simulated  events, we verified that  those backgrounds
are  rendered negligible by the kaon  identification, $\phi$ mass, and
$\Delta   E$  requirements.   The  combinatorial  background  from the
continuum  non-$B \overline  B$  events was estimated using  simulated
events and  the  data collected below $B   \overline B$  threshold. We
found the continuum background to be negligible. 

To estimate the background  contribution from the non-resonant  $B \to
J/\psi \, K^+ \, K^- \, K$ decays, we reconstructed $B^+ \to J/\psi \,
K^+ \,  K^- \,  K^+$  and $B^0 \to   J/\psi  \, K^+ \, K^-   \, K^0_S$
candidates in data requiring  $|M(K^+ K^-)-M_{\phi} |>20$~MeV/$c^2$ to
exclude  $B \to  J/\psi \, \phi   \, K$ events.   We observed 7 $B \to
J/\psi \, K^+ \, K^- \, K$ candidates with  the estimated $B \overline
B$ combinatorial   background of 2.8  events.   We  estimated the mean
background from $B \to J/\psi \, K^+\, K^- \, K$ decays for the $B \to
J/\psi \,  \phi \, K$ signal to   be $0.27^{+0.21}_{-0.17}$ events; we
assumed that  $B \to  J/\psi \,  K^+\, K^- \,  K$ decays  according to
phase space. 

In summary,   the estimated total background for   the combined $B \to
J/\psi \, \phi \, K$ signal is $0.52^{+0.23}_{-0.19}$ events.

We evaluated the reconstruction efficiency using a sample of simulated
$B  \to  J/\psi \,   \phi\, K$ decays.  We  assumed  a uniform  Dalitz
distribution and isotropic decays of $J/\psi$ and $\phi$ mesons; these
assumptions       are              consistent          with       data
(Figs.~\ref{fig:dalitz_data_contour}~and~\ref{fig:paper_polarization_data}).
The reconstruction  efficiency,   which does  not  include   branching
fractions of  daughter  particle decays, is  $(15.5\pm0.2)\%$  for the
$B^+ \to J/\psi \, \phi \, K^+$ mode and $(10.3\pm0.2)\%$ for the $B^0
\to J/\psi \, \phi \, K^0_S$ mode. 
The reconstruction efficiency  is close to zero at  the edges of phase
space where either  $\phi$ or $K$ meson  is produced nearly at rest in
the laboratory frame.  Thus, the overall detection efficiency would be
much smaller than the above values if the $B  \to J/\psi \, \phi \, K$
decay is  dominated  by either a   $J/\psi \,K$ resonance with  a mass
around  4.3~GeV/$c^2$  or a $J/\psi   \, \phi$ resonance   with a mass
around 4.8~GeV/$c^2$. No such resonances  are expected.  To assign the
systematic  uncertainty  due  to  the  decay  model  dependence of the
reconstruction   efficiency, we  generated  two  additional samples of
simulated $B  \to  J/\psi \,   \phi  \, K$  events.   One sample   was
generated with a uniform Dalitz distribution for $B \to J/\psi \, \phi
\, K$  and $100\%$  transverse  polarization for $J/\psi$ and  $\phi$.
The other sample  was generated assuming  the $\phi$ and $K$ mesons to
be daughters  of   a  hypothetical  spin-0  resonance  with mass   
1.7~GeV/$c^2$ and  width 100~MeV.   We estimated the  relative systematic
uncertainty  due to the decay model  dependence  of the reconstruction
efficiency extraction to be 7\%. 

For the branching fraction calculation we  assumed equal production of
$B^+   B^-$ and $B^0  {\overline   B^0}$  pairs at the  $\Upsilon(4S)$
resonance  and ${\cal B}(B^+ \to J/\psi  \,  \phi \, K^+)={\cal B}(B^0
\to J/\psi \, \phi \,  K^0)={\cal B}(B \to  J/\psi \, \phi \, K)$.  We
did not  assign   any   systematic  uncertainty  due   to  these   two
assumptions. We used the world average values of ${\cal B}( J/\psi \to
\ell^+ \ell^-)$, ${\cal B}(  \phi \to K^+  K^-)$, and ${\cal B}( K^0_S
\to    \pi^+   \pi^-)$~\cite{PDG}.   We      used  the   tables     in
Ref.~\cite{Feldman:1998qc} to  assign the 68.27\%  C.L. intervals  for
the Poisson signal mean given the  total number of events observed and
the known mean background.  The resulting branching fraction is ${\cal
B}(B \to J/\psi \, \phi \, K) = (8.8^{+3.5}_{-3.0}[\rm stat]\pm1.3[\rm
syst])\times 10^{-5}$. 

The  systematic error  includes the uncertainty  in the reconstruction
 efficiency due to  decay  modeling plus  the uncertainties  in  track
 finding,   track   fitting, lepton and  charged-kaon  identification,
 $K^0_S$ finding, background subtraction, uncertainty in the number of
 $B \overline B$  pairs used for this  measurement, statistics  of the
 simulated  event  samples,  and  the  uncertainties   on the daughter
 branching fractions ${\cal B}(  J/\psi \to \ell^+ \ell^-)$ and ${\cal
 B}( \phi \to K^+ K^-)$~\cite{PDG}.   We estimated the total  relative
 systematic uncertainty of the  ${\cal B}(B \to J/\psi  \, \phi \, K)$
 measurement to be 15\%. 

In conclusion, we have fully reconstructed 10 $B \to J/\psi \, \phi \,
K$   candidates with a  total   estimated  background  of 0.5  events.
Assuming equal production of $B^+ B^-$ and $B^0 {\overline B^0}$ pairs
at the $\Upsilon(4S)$ resonance  and ${\cal B}(B^+  \to J/\psi \, \phi
\, K^+)={\cal B}(B^0 \to J/\psi \, \phi  \, K^0)={\cal B}(B \to J/\psi
\, \phi \, K)$, we have measured ${\cal B}(B  \to J/\psi \, \phi \, K)
= (8.8^{+3.5}_{-3.0}[\rm   stat]\pm1.3(\rm syst))\times 10^{-5}$. This
is the  first observed $B$  meson decay requiring   the creation of an
additional $s \overline s$ quark pair. 

We gratefully acknowledge the effort of the CESR staff in providing us
with  excellent  luminosity and running   conditions.   This work  was
supported  by the National Science  Foundation, the U.S. Department of
Energy, the   Research Corporation, the  UTPA-Faculty Research Council
Program,  the  Natural Sciences and   Engineering  Research Council of
Canada,  the  A.P.  Sloan   Foundation,   the Swiss National   Science
Foundation, and the Alexander von Humboldt Stiftung.

\begin{thebibliography}{99}
\bibitem{charge-conjugate} 
Charge conjugate modes are implied.  $B \to
J/\psi \,  \phi \, K$ is  either $B^+ \to J/\psi   \, \phi \,  K^+$ or
\linebreak $B^0 \to J/\psi \, \phi \, K^0$.
\end{thebibliography}
\end{document}